\documentstyle[12pt]{article}

\title{
\begin{flushright}
{\large Yaroslavl State University\\
        Preprint YARU-HE-97/05 \\
        hep-ph/9709218} \\[12mm]
\end{flushright}
       {\LARGE\bf Radiative Transition of Massless Neutrino} \\ 
       {\LARGE\bf in Strong Magnetic Field}}

\author{{\Large\bf A.A.~Gvozdev, N.V.~Mikheev} \\[2mm]
        {\large\it Yaroslavl P.G.~Demidov State University, 
                   Yaroslavl 150000, Russia} \\
        {\large\it E-mail: \quad 
                   gvozdev@univ.uniyar.ac.ru \quad 
                   mikheev@yars.free.net} \\[4mm]
        {\Large\bf and} \\[4mm]
        {\Large\bf L.A.~Vassilevskaya} \\[2mm]
        {\large\it Moscow M.V.~Lomonosov State University, 
                   V-952, Moscow 117234, Russia} \\
        {\large\it E-mail: \quad vasilevs@vitep5.itep.ru}}

\date{}

\begin{document}

\maketitle

\bigskip

\begin{abstract}

Radiative transition of massless neutrino
$\nu_l \to \nu_l + \gamma $ ($l = e, \mu, \tau$)
in a strong magnetic field is investigated in the framework 
of the Standard Model. The process probability and the mean values 
of the neutrino energy and momentum loss are presented. 
Possible astrophysical manifestation of the process considered 
is analysed.

\end{abstract}

\vfill

\begin{center}
{\large\it Talk given at the XXXII Rencontres de Moriond} \\
{\large\it on Electroweak interactions and unified theories,} \\
{\large\it Les Arcs, Savoie, France, 15-22 March 1997.}
\end{center}

\thispagestyle{empty}

\newpage

\large

The investigations of neutrino processes in strong magnetic fields
are of principal interest in astrophysics where gigantic neutrino 
fluxes and strong
magnetic fields can take place simultaneously (a process of a 
coalescence of neutron stars, an explosion of a supernova).
By magnetic field is meant the field of strength of order and
more the critical, so called, Schwinger value
$B > B_s$, $B_s = m^{2}_{e}/e \simeq 0.44 \cdot 10^{14} \, G$.
By now astrophysical knowledge of strong magnetic fields
which can be realized in the nature have changed essentially.
For example, the field strengths inside the astrophysical objects
in principle could be as high as $10^{15} - 10^{18} G$ [1].
On the other hand, in such strong magnetic fields 
otherwise negligible processes are not only opened kinematically 
but become substantial ones as well. For example, a photon splitting
into two photons $\gamma \to \gamma + \gamma$ [2], a magnetic
catalysis of the massive neutrino radiative decay [3].

The diagonal radiative neutrino transition becomes possible due 
to the photon dispersion in the strong magnetic field which
differs significantly from the vacuum dispersion 
with increasing photon energy. 
So the real photon  4-momentum can appear as a space-like  
and sufficiently large one ($\vert q^2 \vert \gg m^2_\nu$). 
In this case the phase space for diagonal neutrino transition 
$\nu_l \rightarrow \nu_l + \gamma$ ($l = e, \mu, \tau$) is 
opened also. Thus this diagonal radiative neutrino transition
manifests itself as neutrino bremsstrahlung, does not contain 
uncertainties associated with a possible mixing in the lepton sector 
of the SM and can lead to observable physical effects in the strong 
magnetic fields $B \gg B_s$.

The important property of the amplitude $\nu \to \nu + \gamma$ 
we have studied is that it has a square-root singularities of the type:

\begin{eqnarray}
\sim \left ( {\cal E}^2_{nm} - q^2_{\parallel} \right )^{-1/2} ,
\nonumber
\end{eqnarray}

\noindent in the threshold points 
$(q^2_{\parallel})_{thr} = {\cal E}^2_{nm}$, labelled by $n$, $m$
($n, m = 0,1,2,...$):

\begin{eqnarray}
{\cal E}^2_{nm} 
= \left ( \sqrt{ m^2_e + 2 e B n } + 
 \sqrt{ m^2_e + 2 e B m } \right )^2,  
\nonumber
\end{eqnarray}

\noindent where $q^2_{\parallel} = q^2_0 - q^2_3$ in the frame with
 $\vec B = (0,0,B)$.

\noindent ${\cal E}_{nm}$ has the simple physical meaning
as transversal energy of an electron-positron pair in the 
magnetic field. The similar phenomena is known as the 
cyclotronic resonance in the vacuum polarization [4]. 

It is of interest for some astrophysical applications
the case of relatively high neutrino energy 
$ E_\nu \simeq 10 - 20 MeV \gg m_e$ 
and strong magnetic field $e B \gg E^2_\nu$, when a region 
of the cyclotronic resonance on the ground Landau level 
$q^2_{\parallel} \to 4 m_e^2$ dominates in this process.
The square-root singularities on higher Landau levels are 
not realized under the condition $e B > E^2_\nu$. 

The main contribution to the probability of high energy neutrino
($ e B \gg E_\nu^2 \gg 4 m_e^2$) is determined 
from the vicinity of the resonance point  
$q^2_\parallel = 4 m_e^2$ and can be presented in the form: 

\begin{eqnarray} 
W  \simeq  \frac{\alpha G^2_F}{8 \pi^2 }(g^2_v + g^2_a) 
(e B)^2 E_\nu \sin^2{\theta},
\label{eq:W1} 
\end{eqnarray}

\noindent where $\theta$ is the angle between the vectors 
of the magnetic field strength ${\vec B}$ and the momentum of
the initial neutrino ${\vec p}$;
$g_v$ and $g_a$ are the well-known vector and axial-vector 
$\nu - e$ coupling constants in the SM.

It is of more practical importance for some astrophysical 
applications to calculate the four-vector of the neutrino energy 
and momentum loss in the process of the neutrino radiative 
transition:

\begin{equation}
Q_\mu \, = \, E_\nu \int d W \cdot q_\mu.
\label{eq:Q11}
\end{equation}

\noindent The physical meaning of this vector is that 
$Q_0/E_\nu$ is the mean energy loss in a unit time and
$\vec Q/E_\nu$ is the momentum loss in a unit time.
We have obtained the expression for the vector $Q_\mu$: 
\begin{eqnarray}
Q_\mu & = & \frac{1}{4}\;E_\nu W \Big [ p_\mu -
(p \varphi \varphi)_\mu  -  \frac{2 g_v g_a}{g^2_v + g^2_a} 
(p \tilde\varphi )_\mu \Big ],
\label{eq:Q1} 
\end{eqnarray}

\noindent where $ \varphi_{\alpha \beta} = F_{\alpha \beta} / B$ and 
${\tilde \varphi}_{\alpha \beta} = \frac{1}{2} \varepsilon_{\alpha \beta
\mu \nu} \varphi_{\mu \nu} \; $ are the dimensionless tensor and dual
tensor of the magnetic field.

To illustrate the applications of the results we have obtained
we estimate below possible effects of this process
in astrophysical cataclysm of type of a 
supernova explosion or a merger of neutron stars under the condition
that a compact remnant has for some reasons 
extremely strong magnetic field, 
$B \simeq 10^{16} - 10^{18} G$. 
We assume that neutrinos of all species with the typical mean 
energy $\bar E_\nu \simeq 20 MeV$ 
are radiated from a surface of a neutrinosphere in such a process. 
Using (\ref{eq:Q1})
we estimate the neutrino energy loss:

\begin{eqnarray}
{\cal E}  \simeq  10^{50}\;  
\left (\frac{{\cal E}_{tot}}{ 10^{53} erg} \right )
\left (\frac{B}{10^{17}G} \right )^2
\left (\frac{\bar E_\nu}{20 MeV}\right )
\left (\frac{ R}{10 km}\right ) erg,
\nonumber
\end{eqnarray}

\noindent where ${\cal E}_{tot} \simeq 10^{53}$ erg
is the total neutrino radiation energy;
$B$ is the magnetic field strength in the vicinity of
the neutrinosphere of radius $R$. 
Neutrino bremsstrahlung $\gamma$-quanta are captured 
by a strong magnetic field and propagate along the 
field [4]. Thus the mechanism of
significant power "pumping" of polar caps of a magnetized 
remnant can take place.
This phenomenon can result in reemission within a narrow cone 
$\Omega/4 \pi \ll 1$ along the magnetic moment of the remnant.
So, if an external absorbing envelope is absent in such an
extraordinary astrophysical cataclysm (very strong magnetic fields,
gigantic neutrino fluxes) the reemission process 
will be an observable effect. Namely, such a reemission in the 
magnetic field could appear as an anisotropic $\gamma$-burst with 
the duration of order of the neutrino emission time and of the 
typical energy ${\cal E}_{4\pi}^{burst} \sim 10^{51}$ erg 
in terms of $4 \pi$-geometry.

\vspace{5mm}

The work of N.V.~Mikheev  was supported by a Grant N~d104
from the International Soros Science Education Program.
The work of L.A.~Vassilevskaya  was supported by a fellowship
of INTAS Grant 93-2492-ext and was carried out within the research
program of International Center for Fundamental Physics in Moscow.


\vspace*{-5mm}

\begin{thebibliography}{5}
%
\bibitem{1} 
   M.~Ruderman, 
   in {\it Neutron Stars: Theory and Observation}, 
   ed. by J.~Ventura and D.~Pines, 
   Kluwer Academic. Pub., Dordrecht, 1991; \\
   V.V.~Usov, 
   Nature 357 (1992) 472; \\
   M.~Bocquet at al., 
   Astr. and Ap. 301 (1995) 757.
%
\bibitem{2}
   S.L.~Adler, 
   Ann. Phys. N.Y. 67 (1971) 599. 
%
\bibitem{3}
   A.A.~Gvozdev, N.V.~Mikheev and L.A.~Vassilevskaya, 
   Phys. Rev. D54 (1996) 5674.
%
\bibitem{4}
   A.E.~Shabad, 
   Ann. Phys. N.Y. 90 (1975) 166; \\
   A.E.~Shabad and V.V.~Usov, 
   Astrophysics and Space Science, 102 (1984) 327.
%
\end{thebibliography}
\end{document}